# EL PROBLEMA DE DETERMINAR DEPENDENCIAS FUNCIONALES ENTRE ATRIBUTOS EN LOS ESQUEMAS EN EL MODELO RELACIONAL

## THE PROBLEM DETERMINATION OF FUNCTIONAL DEPENDENCIES BETWEEN ATTRIBUTES RELATION SCHEME IN THE RELATIONAL DATA MODEL

Ignacio Vega-Páez[1], Georgina G. Pulido[2] y José Angel Ortega[3]
ivega@g-ibp.com, gpulido@att.net.mx y oeha430210@hotmail.com
IBP-Memo 2006.07
Sep 2006, México, D.F.

**ABSTRACT**

An alternative definition of the concept is given of "functional dependence" among the attributes of the relational schema in the Relational Model, this definition is obtained in terms of the set theory. For that which a theorem is demonstrated that establishes equivalence and on the basis theorem an algorithm is built for the search of the functional dependences among the attributes. The algorithm is illustrated by a concrete example.

**Terms Key:** Functional Dependence, Data Scheme, BD, database, Relational Model.

**RESUMEN**

Se propone una definición alterna al concepto de "dependencia funcional" entre los atributos de esquemas de relación en el Modelo de Relacional, esta definición se obtiene en términos de la teoría de conjuntos. Para lo cual se demuestra un resultado que establece la equivalencia y con dicho resultado se construye un algoritmo para la búsqueda de las dependencias funcionales entre los atributos de los esquemas involucrados. El algoritmo se ilustra con un ejemplo concreto.

**Términos Clave:** Dependencia Funcional, Esquema de Datos, BD, Base de Datos, Modelo Relacional.

---

[1] Departamento de Computación del Centro de Investigación y de Estudios Avanzados del IPN (CINVESTAV-IPN) y la Dirección de Tecnologías de la Información de la Empresa International Business Partner consulting
[2] Departamento de Ciencias Básicas e Ingeniería de la Universidad Autónoma Metropolitana (UAM).
[3] Sección de Estudios de Posgrado e Investigación de la Escuela Superior de Ingeniería Mecánica y Eléctrica del Instituto Politécnico Nacional (SEPI ESIME IPN)

## INTRODUCCION

Todos los proyectos de desarrollo y construcción de Sistemas de Información Empresariales, son basados en el concepto de Bases de Datos Relacional. En general, por una Base de Datos de un Sistema de Información Empresarial la entenderemos como una colección de datos básicos que están libres de redundancias; mientras que el contenido y la organización de esta colección garantiza una solución efectiva al Sistema de Información Empresarial. En otras palabras, la Base de Datos es un modelo de conocimiento de la Empresa, la cual abstrae las propiedades de la "empresa real" que son importantes desde el punto de vista operacional, administrativo, del negocio, etc.

Actualmente como herramienta formal para la descripción a nivel lógico de una Base de Datos, se usa el Modelo Relacional de Codd [1], este modelo se basa en el concepto de *esquema de relación* ú *objeto relacional*. Este concepto sirve para la descripción de objetos del mismo tipo, y refleja las conexiones entre ellos. Cada una de las propiedades del los objetos son expresadas por medio del concepto de atributo en un esquema de relación (o comúnmente llamado *"relación"*) considerados como conjuntos de valores de dicha propiedad. A cada *atributo* es asignado un único nombre, como regla, se hace que coincida con el nombre de esta propiedad para conservar el significado semántico del atributo. Los elementos del esquema de relación son los *n-tuples* ordenados de valores de los atributos, llamados "*secuencias*", mas formalmente, al conjunto de elementos de un *Esquema de Relación* es un subconjunto del producto cartesiano $A_1 \times A_2 \times ... \times A_n$, donde cada $A_i$ es el *i-ésimo* atributo, que denotaremos $R(A_1, A_2, ..., A_n)$. Al número de atributos de la relación es llamado el grado de la relación.

Cuando se construye una Base de Datos Relacional, los problemas para identificar los elementos en las relaciones y la selección del método de representación dichas relaciones debe ser el más óptimo desde los puntos de vista de la manipulación así como a la no redundancia en el almacenamiento de los datos y ambos son de considerable importancia en la fase de diseño. La solución básica de ambos problemas es el concepto de dependencia funcional entre atributos de los esquemas de relación.

Una dependencia funcional de un atributo $A$ de una relación $R$ sobre un conjunto de atributos $B$ de la misma relación es definida como la dependencia para la cual cada "secuencia" de valores del conjunto de atributos $B$ es asignado para no más de un valor de $A$, que es inducida con

The problem ...

alguna secuencia de $R$. Usaremos la notación tradicional $A \to B$ para denotar la dependencia funcional y denotaremos $A \underset{R}{\to} B$ para dependencia funcional de los atributos $A$ y $B$ de la relación $R$ si hay varias relaciones involucradas.

La presencia de dependencias funcionales entre atributos puede postularse de manera explícita cuando se describe una Base de Datos en la fase de diseño, siguiendo el significado de los datos a modelar (semántica de los objetos a ser descritos). Armstrong [2] investiga posibles familias de tales dependencias funcionales. Sin embargo, el conocimiento incompleto de la semántica de los objetos a ser descritos pueden dejar el problema de dependencias funcionales ocultas. El problema de descubrir tales dependencias funcionales en una base de datos real es el problema que se plantea en este trabajo, además que es un problema latente que "normalmente" aparece hasta la entrada en producción. El algoritmo que proponemos al final es la solución a este problema.

**DEFINICION Y PLANTEAMIENTO DEL PROBLEMA**

Consideremos una relación $R(A_1, A_2,..., A_n)$, donde $A_i$ son los atributos de la relación. Introduzcamos un atributo $K$ que conste de un número ordinal con el rol de la secuencia que enumera todos los elementos de la relación $R$, que llamaremos *"enumeración"* de $R$ esta obviamente, define unívocamente a cada una de las secuencias de esta relación,

$$R(K, A_1, A_2,..., A_n). \tag{1}$$

La proyección de este conjunto de secuencias de la relación $R$ sobre los $K$ y $A_i$, formalmente $p_{\{K, A_i\}}(R)$ (Ullman [3]), puede ser escrito de la forma,

$$F_i = \{\langle k, a_i \rangle | k \in K, a_i \in A_i\}. \tag{2}$$

Entonces el triplete $\langle F_i, K, A_i \rangle$ define a la función $f_i$ para toda $i \in [\![1, n]\!]$, que satisface las siguientes propiedades:

a) $F_i \subseteq K \times A_i$

b) En vista del uni-evaluación de la enumeración (1), y en (2) no hay ningún par con el primer elemento repetido

$$f_i \underset{Df}{=} \langle F_i, K, A_i \rangle \quad \text{ó} \quad A_i = f_i(K) \tag{3}$$

La función es suprayectiva ó sobre, porque se cumple:

$$\left(\forall a_i^k \in A_i\right)\left(\exists k \in K\right)\left[f_i(k) = a_i^k\right] \tag{4}$$

y por consiguiente el sistema de clases $\overline{K}_i$ generado por $f_i^{-1}(A_i) = \overline{K}_i$ es una partición de del conjunto $K$.

Si de notamos $f_i^{-1}$ por $\boldsymbol{f}_i$, podemos escribir

$$\overline{K}_i = \Phi_i(A_i) \tag{5}$$

## RESULTADOS

**PROPOSICION:** Para cada $i$, La función $\boldsymbol{f}_i$ es biyectiva.

En efecto: para cada $a_i^m \in A_i$ es identificado con solo una clase $\overline{K}_i^m \in \overline{K}_i$, y diferentes clases $\overline{K}_i^{m_1}$ y $\overline{K}_i^{m_2}$ son identificados por diferentes $a_i^{m_1}$ y $a_i^{m_2}$; cada clase $\overline{K}_i^m$ es identificado con solo un $a_i^m$.

El siguiente teorema usa las construcciones anteriores (1)-(5).

**TEOREMA** (Dependencia Funcional Conjuntista). Un atributo $A_{i_2}$ depende funcionalmente de un atributo $A_{i_1}$ si y sólo si cada clase de la partición $\overline{K}_{i_1}$ es subconjunto de al menos una clase de la partición $\overline{K}_{i_2}$, ó formalmente

$$\left(\forall m, \exists n, \overline{K}_{i_1}^m \subseteq \overline{K}_{i_2}^n\right) \Leftrightarrow A_{i_1} \underset{R}{\to} A_{i_2} \tag{6}$$

**Demostración:**

**La necesidad:** En base a (5) para $i_1$ e $i_2$ se tiene

$$\overline{K}_{i_1} = \Phi_{i_1}(A_{i_1}), \ \overline{K}_{i_2} = \Phi_{i_2}(A_{i_2}) \tag{7}$$

The problem ...

Por la biyectividad de las funciones $f_i$

$$A_{i_2} = \Phi_{i_2}^{-1}(\overline{K}_{i_2}) \tag{8}$$

se tiene $\forall m, \exists n, \overline{K}_{i_1}^m \subseteq \overline{K}_{i_2}^n$, por lo que existe una función $j$ sobre, tal que

$$\overline{K}_{i_2} = j\left(\overline{K}_{i_1}\right) \tag{9}$$

entonces en base a las expresiones de 7-9 podemos escribir

$$A_{i_2} = \Phi_{i_2}^{-1}(\overline{K}_{i_2}) = \Phi_{i_2}^{-1}(j(\overline{K}_{i_1})) = \Phi_{i_2}^{-1}(j(\Phi_{i_1}(A_{i_1}))) = F(A_{i_1})$$

Por la composición de funciones es una función, $A_{i_2}$ depende funcionalmente de $A_{i_1}$

**La suficiencia:** por reducción al absurdo, demostraremos que si,

$$\exists m, \forall n, \overline{K}_{i_1}^m \nsubseteq \overline{K}_{i_2}^n \tag{10}$$

entonces $A_{i_1} \not\to_R A_{i_2}$.

De la condición (9) se sigue que existen $k'$ y $k'' \in \overline{K}_{i_1}^{m_1}$ tal que $k' \in \overline{K}_{i_2}^{n_1}$ mientras $k'' \in \overline{K}_{i_2}^{n_2}$

Esto es consecuencia de que la partición $\overline{K}_{i_1}$, $k'$ y $k''$ pertenecen a una sola clase $\overline{K}_{i_1}^{m_1}$, y por la biyectividad de $f_{i_1}$ (ver (5)) un solo valor $a_{i_1}^{m_1}$ corresponde a ambos. En la partición $\overline{K}_{i_2}$ $k'$ esta en la clase $\overline{K}_{i_2}^{n_1}$ y le corresponde el valor $a_{i_2}^{n_1}$, mientras $k''$ esta en la clase $\overline{K}_{i_2}^{n_2}$ y le corresponde el valor $a_{i_2}^{n_2}$. En otros palabras, uno y el mismo valor del atributo $A_{i_1}$, en diferentes enumeraciones del esquema de relación son aparejados por diferentes valores del atributo $A_{i_2}$, lo cual contradice la definición de dependencia funcional. ¦

**OBSERVACION:** Si el conjunto de clases de la partición corresponden al conjunto de valores de la colección de atributos $A_{j_1}, A_{j_2}, ..., A_{j_x}$ es definido como

$$\overline{K}_{j_1, j_2, ..., j_x} = \overline{K}_{j_1} \cap \overline{K}_{j_2} \cap ... \cap \overline{K}_{j_x}$$

entonces para la dependencia funcional del atributo $A_i$ sobre la colección de atributos $A_{j_1}, A_{j_2},..., A_{j_x}$ el teorema se toma la forma siguiente

$$\left(\forall m, \exists n, \overline{K}^{m}_{j_1,j_2,...,j_x} \subseteq \overline{K}^{n}_{i}\right) \Leftrightarrow \left(A_{j_1}, A_{j_2},..., A_{j_x}\right) \xrightarrow[R]{} A_i \qquad (11)$$

**COROLARIO 1**.

$$\overline{K}_{i_1} = \overline{K}_{i_2} \Leftrightarrow A_{i_1} \underset{R}{\leftrightarrow} A_{i_2}$$

El corolario se demuestra en base a una substitución simétrica de los índices $i_1$ e $i_2$ en el teorema.

Para identificar un elemento de la relación podemos usar cualquier atributo o una colección de atributos sobre cualquiera otros atributos de la relación de los cuales dependen funcionalmente. Tal atributo o colección de atributos (de los cuales no podemos quitar uno de los atributos, sin perturbar esta dependencia) se llama un *llave candidata*. Los valores de los atributos de cualquier llave candidata identifican unívocamente a los elementos de la relación.

**COROLARIO 2** Un conjunto mínimo $A_{j_1}, A_{j_2},..., A_{j_x}$ al cual el conjunto de clases de la partición

$$\overline{K}_{j_1} \cap \overline{K}_{j_2} \cap ... \cap \overline{K}_{j_x} = \{\{1\},\{2\},...,\{p\}\},$$

que corresponda, es llave candidata de la relación.

**ALGORITMO**

Por construcción el atributo $K$ es una llave candidata de la relación. La partición del conjunto $K$ para este atributo es el conjunto

$$\{\{1\},\{2\},...,\{p\}\}$$

Entre las "llaves candidatas" de una relación existe una dependencia funcional mutua; por consiguiente en base a "Corolario 1" las particiones del conjunto $K$ corresponden a éstas colecciones de atributos que deben coincidir.

Con base al teorema demostrado proponemos el siguiente algoritmo para determinar la dependencia funcional entre atributos de una relación.

The problem ...

1. Para cada atributo de la relación construimos una proyección léxicográficamente ordenada del conjunto de enumeraciones de la relación $R$ sobre $K$ y $A_i$ $(i = 1,2,...n)$.

   Como resultado de la construcción de los pares, el conjunto de los segundos elementos donde los primeros elementos son iguales para cada proyección forman la partición del conjunto $K$ para los atributos correspondientes.

2. Para determinar la presencia de una dependencia funcional entre dos atributos de la relación hay que verificar la intersección de cada clase de la partición generada por un atributo con las clases de la partición generada por el otro atributo. Si al menos una clase de la partición del primer atributo no es subconjunto de la misma clase de partición del segundo atributo, podemos concluir la ausencia de una dependencia funcional sobre el primer atributo (por la necesidad de nuestro teorema, ver expresión (6)).

   Cuando la clase del primer atributo, que consistente en un solo elemento, siempre es un subconjunto de alguna clase de la partición del segundo atributo, es aconsejable sólo llevar a cabo una verificación para las clases que contienen más de dos elementos.

   Hay otra afirmación obvia, para reducir el número de comparaciones, que consiste del hecho que tenemos que comparar sólo esas clases del segundo atributo para el cual el número de elementos no es menor al número de elementos de la clase del primer atributo a ser comparado.

Por medio del algoritmo presentado podemos encontrar todas las dependencias funcionales de la relación. Para esto, se sigue de (11), debemos construir todos los conjuntos de las clases de la partición correspondientes a los conjuntos de valores de todas las colecciones de atributos $A_{j_1}, A_{j_2},..., A_{j_x}$.

**EJEMPLO**

**Ejemplo.** Tomemos la relación $EnvioPostal(Clave, Color, Volumen, Peso)$

Apliquemos la enumeración $K$ a la relación $EnvioPostal$

$$EnvioPostal'(K, Clave, Color, Volumen, Peso)$$

La tabla siguiente muestra lo anterior:

| K | CLAVE | COLOR | VOLUMEN | PESO |
|---|---|---|---|---|
| 1 | A1 | ROJO | 15 | 150 |
| 2 | B2 | AZUL | 20 | 230 |
| 3 | CA1 | AMARILLO | 18 | 160 |
| 4 | CB2 | VERDE | 40 | 420 |
| 5 | C4 | AMARILLO | 18 | 160 |
| 6 | 3 | AZUL | 25 | 210 |
| 7 | 4 | VERDE | 40 | 360 |
| 8 | 5 | NEGRO | 60 | 540 |

Las particiones siguientes del conjunto $K$ corresponden a los atributos de la relación *EnvioPostal'* representados en la tabla anterior.

$$\overline{K}_{Clave} = \{\{1\},\{2\},\{3\},\{4\},\{5\},\{6\},\{7\},\{8\}\}$$

$$\overline{K}_{Color} = \{\{1\},\{2,6\},\{3,5\},\{4,7\},\{8\}\}$$

$$\overline{K}_{Volumen} = \{\{1\},\{2\},\{3,5\},\{4,7\},\{6\},\{8\}\}$$

$$\overline{K}_{Peso} = \{\{1\},\{2\},\{3,5\},\{4\},\{6\},\{7\},\{8\}\}$$

En Tabla siguiente:

| ATRIBUTOS | CLAVE | COLOR | VOLUMEN | PESO |
|---|---|---|---|---|
| CLAVE | 1 | 0 | 0 | 0 |
| COLOR | 1 | 1 | 1 | 1 |
| VOLUMEN | 1 | 0 | 1 | 1 |
| PESO | 1 | 0 | 0 | 1 |

el valor 1 indica la presencia de una dependencia funcional de la fila $n$ de la columna correspondiente. Por lo que el rol de llave de la relación la podemos verificar ó escoger al atributo "CLAVE"

## CONCLUSION

El algoritmo presentado nos permite verificar la presencia o ausencia de una dependencia funcional entre los atributos de una relación para una enumeración. Es obvio entre más grande el número de enumeraciones a ser analizadas, será mayor la certeza de la conclusión a que se llegue. La pregunta final sobre la presencia de una dependencia funcional entre los atributos de una relación de un conjunto arbitrario de enumeraciones puede resolverse mediante un análisis

The problem ...

de la semántica de los atributos o simplemente verificar la semántica de los atributos con las dependencias funcionales intuidas en el diseño.

**REFERENCIAS**

1. E. F. Codd, "A relational model of data for large shared data banks", Commun. ACM, 13, No. 6, 377 (1970).
2. W. W. Armstrong, "Dependency structures of database relationships", Proc. IFIP Congr.3, 580 (1974)
3. Jeffrey D. Ullman, "Principles of Database Systems", Computer Science Press, Inc. (1980)
4. Vega-Páez I. y F.D. Sagols T., *"Space Program Language (SPL/SQL) for Relational Approach of the Spatial Databases",* Proceedings in Information Systems Analysis and Synthesis ISAS '95, 5th International Symposium on Systems Research, Informatics and Cybernetics, pp. 89-93 August 16-20, 95, Baden-Baden, Germany (1995)